\documentclass[%
 reprint,
superscriptaddress,
showpacs,preprintnumbers,
 amsmath,amssymb,
 aps,prl
]{revtex4-1}

\usepackage{graphicx}% Include figure files
\usepackage{dcolumn}% Align table columns on decimal point
\usepackage{bm}% bold math
\DeclareMathOperator{\Tr}{Tr}

\begin{document}

\preprint{APS/123-QED}

\title{Exchange interactions in the Hubbard-Stratonovich transformation for the stability analysis of itinerant ferromagnetism}
%\title{Stability analysis of itinerant ferromagnetism for a spin-1/2 Fermi gas with a general interaction potential}

\author{Enya Vermeyen}
\email{enya.vermeyen@uantwerpen.be}
\affiliation{TQC, Universiteit Antwerpen, B-2610 Antwerpen, Belgium}

\author{Carlos A.R. S\'{a} de Melo}
\affiliation{School of Physics, Georgia Institute of Technology, Atlanta 30332, USA}

\author{Jacques Tempere}
\affiliation{TQC, Universiteit Antwerpen, B-2610 Antwerpen, Belgium}
\affiliation{Lyman Laboratory of Physics, Harvard University, Cambridge, Massachusetts 02138, USA}

\date{November 14, 2015}% It is always \today, today,

\begin{abstract}
Itinerant ferromagnetism, i.e. spontaneous polarization of non-localized particles, is expected to occur for strong repulsive interactions in a spin-1/2 Fermi system. However, this state has proven notoriously hard to find experimentally, both in ultracold gases and in solids. This raises questions about the stability of the itinerant ferromagnetic state itself. Here we develop a new approach to describe both the direct and exchange interactions for a general interaction potential in the path-integral formalism and we apply this method to itinerant ferromagnetism in three-dimensional ultracold Fermi gases.
We show that the exchange interactions are lost in the Hubbard-Stratonovich transformation and we propose to explicitly include the exchange effects in a new modified interaction potential. In the saddle-point approximation, the effect of interactions can be taken into account using only three parameters. If the interactions become too strong, all saddle points become unstable to density fluctuations. This greatly restricts the area in the phase diagram where uniform itinerant ferromagnetism is expected to occur.
%\begin{description}
%\end{description}
\end{abstract}

\maketitle

%\tableofcontents

\textit{Introduction} In 1929, Bloch suggested that the exchange energy may overcome the kinetic energy cost of polarization in a (non-localized) gas of spin 1/2 fermions with strong repulsive interactions \cite{Bloch}, a phenomenon which is called itinerant ferromagnetism. Despite this long history and its importance for d-band transition metals \cite{Stoner,dband}, itinerant ferromagnetism remains poorly understood. This is due to two factors. First, the strong interactions and correlations involved make it difficult to create accurate theoretical models. Second, so far itinerant ferromagnetism has not yet been observed in its pure form \cite{dband}, making it difficult to verify and improve theoretical predictions.

Due to their great experimental tunability, ultracold atomic gases have been proposed as a model system for the experimental realization of itinerant ferromagnetism  \cite{predictioninuc,prediction1,prediction2,prediction3,duineprediction}. Using Feshbach resonances, the interaction strength can be tuned, enabling the realization of the strong repulsive interactions where itinerant ferromagnetism is expected to occur.
In a recent experiment, the Ketterle group was able to probe this regime \cite{mitexp1,mitexp1analysis}, but fast molecular pairing due to the instability of the repulsive branch of the Feshbach resonance prevented the formation of any equilibrium state (precluding formation of the itinerant ferromangetic state) \cite{mitexp2proposal,mitexp2,mitexptheory1,mitexptheory2}. In response, there have been several proposals for minimizing the effects of this experimental instability \cite{mixtureproposal,socproposal1,*socproposal2,1dproposal1,%
1dproposal2,latticeproposal1,latticeproposal2,fluxlatticeproposal,%
honeycombproposal}.

Inspired by these experimental advances, we have revisited the theory of itinerant ferromagnetism. In our earlier work \cite{contactinstability}, we already studied this problem for contact interactions in the path integral approach and we found that the itinerant ferromagnetic state is unstable within the saddle-point approximation. However, it was not clear if and under which conditions the itinerant ferromagnetic state can be stabilised for other interaction potentials \cite{stability1,stability2,stability3,stability4}. This question is especially relevant in light of the new interaction potentials (most notably dipolar interactions \cite{dipolar1,*dipolar2,*dipolar3,%
dipolar4,dipolar5,dipolar6,dipolar7}, but also p-wave interactions \cite{feshbachreview}), which are being probed in ultracold atomic Fermi gases.

In this Letter, we introduce a new framework for describing the direct and exchange interactions for a general interaction potential in the path-integral formalism and we apply this method to itinerant ferromagnetism in three-dimensional (3D) ultracold atomic gases.

\textit{Formalism} We calculate the thermodynamic grand potential per unit volume $\Omega=\ln\left(\mathcal{Z}\right)/\beta V$ of a (pseudo)spin-1/2 Fermi gas with $\mathcal{Z}$ the partition sum, $\beta =1/k_{B}T$ the inverse temperature and $V$ the volume. In the path-integral formalism, the partition sum can be calculated by summing over all possible configurations of the fermionic Grassmann fields $\psi _{\uparrow}$ and $\psi _{\downarrow}$ (and their conjugated counterparts $\bar{\psi}_{\uparrow}$ and $\bar{\psi}_{\downarrow}$), weighted by the action $S$ of each configuration: $\mathcal{Z}=\prod_{\sigma =\uparrow,\downarrow}\int \mathcal{D}\bar{\psi}_{\sigma}\int\mathcal{D}\psi_{\sigma}\exp\left(-S\left[\bar{\psi},\psi\right]\right) $. The action of the system (in units $\hbar=1$, $k_{B}=1$, the mass of the particles $m=1/2$ and the Fermi wave vector $k_{F}=1$) is given by%
\begin{eqnarray}
&&S\left[  \bar{\psi},\psi \right] =\sum_{\sigma_{1}}\int \limits_{0}^{\beta}d\tau \int \limits_{V}d\mathbf{x}\label{action}\\
&&\left \{  \bar{\psi}_{\sigma_{1},\mathbf{x},\tau}\left(  \frac{\partial}{\partial \tau}-\mathbf{\nabla}_{\mathbf{x}}^{2}-\mu_{\sigma_{1}}\right)\psi_{\sigma_{1},\mathbf{x},\tau}\right.\notag\\
&&\left. +\sum_{\sigma_{2}}\int \limits_{V}d\mathbf{x}^{\prime}\frac{g_{\sigma_{1}\sigma_{2}}\left(  \mathbf{\Delta x}\right)}{2}  \bar{\psi}_{\sigma_{1},\mathbf{x},\tau}\psi_{\sigma_{1},\mathbf{x},\tau}\bar{\psi}_{\sigma_{2},\mathbf{x}^{\prime},\tau}\psi_{\sigma_{2},\mathbf{x}^{\prime},\tau}\right \}  \text{,}\notag
\end{eqnarray}%
with $\tau$ the imaginary time, $\mu_{\sigma}$ the spin $\sigma$ chemical potentials, $g_{\sigma_{1}\sigma_{2}}\left(\mathbf{\Delta x}\right)$ the interaction potential and $\mathbf{\Delta x}=\mathbf{x-x}^{\prime}$. For symmetry reasons, we will assume $g_{\uparrow\downarrow}\left(\mathbf{\Delta x}\right)=g_{\downarrow\uparrow}\left(-\mathbf{\Delta x}\right)$.

Due to the presence of the interaction term in the action, the path integral cannot be calculated exactly for a general case. Instead, the interaction term is decoupled into several terms of second order in the fermionic fields by introducing an auxiliary bosonic field through the Hubbard-Stratonovich transformation. We will choose to use the real density fields $\rho_{\uparrow}$ and $\rho_{\downarrow}$ from the Hartree channel as bosonic auxiliary fields \cite{HSKleinert}. The transformation is constructed by shifting the real bosonic fields $\rho_{\uparrow}$ and $\rho_{\downarrow}$ by the reference fields $\rho_{\uparrow}^{0}$ and $\rho_{\downarrow}^{0}$ in the path integral%
\begin{widetext}
\begin{equation}
\mathcal{Z}_{\rho } =\int \mathcal{D\rho }_{\uparrow }\int \mathcal{D\rho }_{\downarrow }\exp\left \{ \frac{1}{2}\sum_{\sigma _{1},\sigma _{2}=\uparrow ,\downarrow}\int\limits_{0}^{\beta }d\tau \int\limits_{V}d\mathbf{x}\int\limits_{V}d\mathbf{x}^{\prime}g_{\sigma _{1}\sigma _{2}}\left( \mathbf{\Delta x}\right) \left[ \rho_{\sigma _{1},\mathbf{x},\tau} -\rho _{\sigma _{1},\mathbf{x},\tau}^{0} \right] \left[ \rho _{\sigma _{2},\mathbf{x}^{\prime},\tau} -\rho _{\sigma _{2},\mathbf{x}^{\prime},\tau}^{0} \right] \right \} \text{,}
\end{equation}
\end{widetext}%
after which the term $\rho_{\sigma_{1},\mathbf{x},\tau}^{0}\rho_{\sigma _{2},\mathbf{x}^{\prime},\tau}^{0}$ is isolated and chosen to be equal to $\bar{\psi}_{\sigma_{1},\mathbf{x},\tau}\psi _{\sigma_{1},\mathbf{x},\tau}\bar{\psi}_{\sigma_{2},\mathbf{x}^{\prime},\tau}\psi_{\sigma_{2},\mathbf{x}^{\prime},\tau} $. In the case of the Hartree channel, the reference fields correspond to the fermionic density: $\rho_{\sigma,\mathbf{x},\tau}^{0}=\bar{\psi}_{\sigma,\mathbf{x},\tau}\psi _{\sigma,\mathbf{x},\tau}$. This results in the following transformation,%
\begin{widetext}
\begin{eqnarray}
&&\exp \left[ -\frac{1}{2}\sum_{\sigma _{1},\sigma _{2}=\uparrow ,\downarrow}\int\limits_{0}^{\beta }d\tau \int\limits_{V}d\mathbf{x}\int\limits_{V}d\mathbf{x}^{\prime}g_{\sigma _{1}\sigma _{2}}\left( \mathbf{\Delta x}\right) \bar{\psi}_{\sigma _{1},\mathbf{x},\tau} \psi _{\sigma _{1},\mathbf{x},\tau} \bar{\psi}_{\sigma _{2},\mathbf{x}^{\prime},\tau} \psi _{\sigma _{2},\mathbf{x}^{\prime},\tau}\right] = \frac{1}{\mathcal{Z}_{\rho }}\int \mathcal{D\rho }_{\uparrow }\int \mathcal{D\rho }_{\downarrow }\label{transformation} \\
&&\times \exp \left \{ \frac{1}{2}\sum_{\sigma _{1},\sigma _{2}=\uparrow,\downarrow }\int\limits_{0}^{\beta }d\tau \int\limits_{V}d\mathbf{x}\int\limits_{V}d\mathbf{x}^{\prime }g_{\sigma _{1}\sigma _{2}}\left( \mathbf{\Delta x}\right)\left[ \rho _{\sigma _{1},\mathbf{x},\tau} \rho_{\sigma _{2},\mathbf{x}^{\prime },\tau} -\rho _{\sigma_{1},\mathbf{x},\tau} \bar{\psi}_{\sigma _{2},\mathbf{x}^{\prime },\tau} \psi _{\sigma _{2},\mathbf{x}^{\prime },\tau} -\bar{\psi}_{\sigma _{1},\mathbf{x},\tau} \psi_{\sigma _{1},\mathbf{x},\tau} \rho _{\sigma _{2},\mathbf{x}^{\prime },\tau} \right] \right \} \text{.}\notag
\end{eqnarray}
\end{widetext}%
The prefactor $\mathcal{Z}_{\rho}$ shifts the zero point of the thermodynamic grand potential and will be taken as our energy reference. In the left-hand side of eq. (\ref{transformation}), the product of two equal Grassmann variables is zero, ensuring that interactions with $\mathbf{x=x}^{\prime}$ and $\sigma_{1}=\sigma_{2}$ are not taken into account. However, in the right-hand side there is no similar mechanism to take into account the exchange interactions. This discrepancy is caused by the fact that the new reference field $\bar{\psi}_{\sigma,\mathbf{x},\tau}\psi _{\sigma,\mathbf{x},\tau}$ is  a product of Grassmann variables and not a complex number, while bosonic fields should be complex. The fermionic symmetry properties of a pair of Grassmann variables are lost when treating that pair as a complex bosonic field, an effect which becomes important when studying the exchange interactions.

In order to include the exchange interactions, we propose to explicitly enforce the Pauli principle using a modified interaction potential $\tilde{g}_{\sigma_{1}\sigma_{2}}\left(\mathbf{\Delta x}\right)=g_{\sigma_{1}\sigma_{2}}\left(\mathbf{\Delta x}\right)\left[1-f_{\sigma_{1}}\left(\mathbf{\Delta x}\right)\delta _{\sigma_{1}\sigma _{2}}\right]$ with $\delta_{\sigma_{1}\sigma_{2}}$ the Kronecker delta and $f_{\sigma}\left(\mathbf{\Delta x}\right)$ a ``shielding" function. Interaction processes with $\mathbf{x=x}^{\prime}$ and $\sigma_{1}=\sigma_{2}$ are excluded by choosing $f_{\sigma}\left(\mathbf{\Delta x}\right)=\delta_{1}\left(\mathbf{\Delta x}\right) $, where $\delta_{1}\left(\mathbf{\Delta x}=\mathbf{0}\right)=1$ and $\delta_{1}\left( \mathbf{\Delta x}\neq\mathbf{0}\right)=0$. This discrete choice of the shielding function only makes sense when considering a countable basis of the Hilbert space. The position basis is an uncountable set, which is often used for practical purposes as the unphysical continuum limit of real physical countable bases. In order to compensate for the difference, the shielding function $f_{\sigma}\left(\mathbf{\Delta x}\right)$ should be modified in such a way that a certain (small) range of interactions is excluded. In experiments only the modified interaction (pseudo)potential can be measured: the real interaction (pseudo)potential $g_{\sigma_{1}\sigma_{2}}\left(\mathbf{\Delta x}\right)$ and the shielding $f_{\sigma _{1}}\left(\mathbf{\Delta x}\right)$ of interactions between particles of the same spin state remain unknown, although they could be modeled.

After the Hubbard-Stratonovich transformation, a new effective action is defined, which is Fourier transformed in order to remove the derivatives from the kinetic energy. If $\tilde{g}_{\sigma_{1}\sigma_{2}}\left(\mathbf{\Delta x}\right)=\tilde{g}_{\sigma_{1}\sigma_{2}}\left(-\mathbf{\Delta x}\right)$, one obtains%
\begin{eqnarray}
S_{\rm{eff}}&&\left[ \bar{\psi},\psi ,\rho \right] =\sum_{\sigma _{1}=\uparrow,\downarrow }\sum \limits_{\mathbf{k},n}\sum \limits_{\mathbf{k}^{\prime},n^{\prime }}\bar{\psi}_{\sigma _{1},k} \left[ -G_{\sigma_{1}}^{-1}\left( k,k^{\prime }\right) \right] \psi _{\sigma _{1},k^{\prime}}  \notag \\
&&-\frac{\sqrt{V}}{2}\sum_{\sigma _{1},\sigma _{2}=\uparrow ,\downarrow}\sum \limits_{\mathbf{Q},m}\tilde{g}_{\sigma _{1}\sigma _{2}}\left( \mathbf{Q}\right) \rho _{\sigma _{1},-Q} \rho _{\sigma _{2},Q} \text{,}
\end{eqnarray}%
with the four-vectors $k=\left(\mathbf{k},\omega_{n}\right)$ and $Q=\left(\mathbf{Q},\Omega_{m}\right)$, $\omega_{n}$ the fermionic and $\Omega_{m}$ the bosonic Matsubara frequencies,
\begin{eqnarray}
-G_{\sigma _{1}}^{-1}\left( k,k^{\prime }\right) &=&\left( -i\omega _{n}+\mathbf{k}^{2}-\mu _{\sigma _{1}}\right) \delta \left( \Delta k\right)\notag \\
&+& \frac{1}{\sqrt{\beta }}\sum_{\sigma _{2}=\uparrow ,\downarrow }\tilde{g}_{\sigma _{1}\sigma _{2}}\left( \Delta k\right) \rho _{\sigma _{2},\Delta k}
\end{eqnarray}%
the inverse Green's function, $\delta\left(\Delta k\right)$ the Dirac delta function and $\Delta k=k-k^{\prime}$. After performing the fermionic path integral, the partition sum is given by%
\begin{widetext}
\begin{equation}
\mathcal{Z}=\int \mathcal{D\rho }_{\uparrow }\int \mathcal{D\rho }_{\downarrow }\exp\left( \frac{\sqrt{V}}{2}\sum_{\sigma _{1},\sigma _{2}=\uparrow ,\downarrow}\sum \limits_{\mathbf{Q},m}\tilde{g}_{\sigma _{1}\sigma _{2}}\left( \mathbf{Q}\right) \rho _{\sigma _{1},-Q} \rho _{\sigma _{2},Q} +\sum_{\sigma _{1}=\uparrow ,\downarrow }\Tr\left \{ \ln \left[-G_{\sigma _{1}}^{-1}\left( k,k^{\prime }\right) \right] \right \} \right) \text{.}  \label{zrho}
\end{equation}
\end{widetext}

\textit{Saddle-point approximation} The remaining bosonic path integral in eq. (\ref{zrho}) cannot be calculated exactly for a general case. In the saddle-point approximation, the densities are assumed to be constant: $\rho_{\sigma,Q}=\sqrt{\beta V}\delta\left(Q\right) \rho_{\sigma }$. This results in the following expression for the saddle-point thermodynamic grand potential as a function of $\left( \beta ,\mu_{\uparrow},\mu_{\downarrow};\rho_{\uparrow},\rho_{\downarrow }\right)$ in D dimensions,%
\begin{equation}
\Omega_{sp} =-\frac{1}{2}\sum_{\sigma_{1},\sigma_{2}=\uparrow,\downarrow}\tilde{g}_{\sigma_{1}\sigma_{2}}\rho_{\sigma_{1}}\rho_{\sigma_{2}}+\Omega_{sp,kin}\left(\beta,\mu_{\uparrow}^{\prime},\mu_{\downarrow}^{\prime}\right)\text{,}
\label{spenergy}
\end{equation}%
where we defined new interaction parameters $\tilde{g}_{\sigma _{1}\sigma_{2}}=\sqrt{V}\tilde{g}_{\sigma _{1}\sigma _{2}}\left( \mathbf{0}\right) $ and an effective chemical potential 
\begin{equation}
\mu _{\sigma _{1}}^{\prime }=\mu _{\sigma _{1}}-\sum_{\sigma _{2}=\uparrow,\downarrow }\tilde{g}_{\sigma _{1}\sigma _{2}}\rho _{\sigma _{2}}\text{.}\label{effchempoteq}
\end{equation}%
The first term in eq. (\ref{spenergy}) represents the interaction energy, while the second one represents the kinetic energy,
\begin{eqnarray}
&&\Omega _{sp,kin}\left( \beta ,\mu _{\uparrow }^{\prime },\mu _{\downarrow}^{\prime }\right)= -\sum_{\sigma _{1}=\uparrow ,\downarrow }\int \frac{d^{D}k}{\left( 2\pi \right) ^{D}}\label{kineticenergy}\\ 
&&\times\left( \frac{1}{\beta }\ln \left \{ 1+\exp \left[ -\beta \left( \mathbf{k}^{2}-\mu _{\sigma _{1}}^{\prime }\right) \right] \right \} -\mathbf{k}^{2}+\mu _{\sigma _{1}}^{\prime }\right) \text{.}\notag
\end{eqnarray}%
It has the same form as the kinetic energy of the non-interacting gas, but its chemical potentials are shifted by the interactions.

In eqs. (\ref{spenergy}) and (\ref{kineticenergy}), the values of $\rho_{\uparrow}$ and $\rho_{\downarrow}$ still have to be determined using the saddle-point equations, $\left.\partial\Omega_{sp}\left( \beta,\mu_{\uparrow},\mu_{\downarrow};\rho_{\uparrow},\rho_{\downarrow}\right)/\partial\rho_{\sigma}\right\vert_{\beta,\mu_{\uparrow},\mu _{\downarrow};\rho_{-\sigma}}=0$. The particle number density of each spin state $\sigma$ is given by the number equations $n_{\sigma}=-\left.\partial\Omega_{sp}\left(\beta,\mu_{\uparrow},\mu _{\downarrow}\right)/\partial\mu _{\sigma}\right\vert_{\beta,\mu_{-\sigma}}$. Because we use $k_{F}=\sqrt[3]{3\pi ^{2}n}$ as a unit with $n=n_{\uparrow}+n_{\downarrow}$, the saddle-point equations have to be solved together with the number equation $n=1/3\pi ^{2}$. The saddle-point equations can be rewritten to find $\rho_{\sigma}=n_{\sigma}$.

A solution to the saddle-point equations can only be stable against density fluctuations if it is also a minimum of $\Omega_{sp}$ as a function of $\rho_{\uparrow}$ and $\rho_{\downarrow}$. This is the case if the Hessian matrix $H$ of second derivatives of $\Omega_{sp}$ to $\rho_{\uparrow}$ and $\rho_{\downarrow}$ has two positive eigenvalues, i.e. if both its trace and determinant are positive.

\begin{figure}
\includegraphics[width=\linewidth]{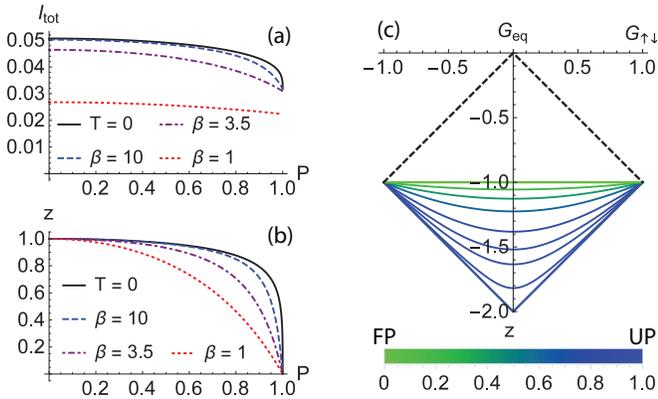}% Here is how to import EPS art
\caption{\label{stabilityfigures} (a) and (b): The parameters $I_{tot}$ (a) and $z$ (b) as a function of the polarization $P$, for different values of the inverse temperature: $\beta\rightarrow+\infty $ ($T=0$, black solid line), $\beta=10$ (blue dashed line), $\beta=3.5$ (purple dot-dashed line) and $\beta=1$ (red dotted line). (c) The stability areas for different values of $z$ as a function of the scaled interaction parameters $G_{\uparrow\downarrow}=I_{tot}\tilde{g}_{\uparrow\downarrow}$ and $G_{eq}=I_{tot}\tilde{g}_{eq}$. The black dashed line is the upper boundary of the stability area (independent of $z$), while the colored solid lines are the lower boundaries of the stability area for different values of $z$.}
\end{figure}

\textit{Example} In order to study itinerant ferromagnetism in ultracold atomic gases, we will consider the case $\tilde{g}_{\downarrow\downarrow}=\tilde{g}_{\uparrow\uparrow}=\tilde{g}_{eq}$ and $\mu_{\uparrow}=\mu_{\downarrow}=\mu$ in 3D. In that case, polarization can only be induced by the interactions. We construct a phase diagram which shows as a function of the interaction parameters $\tilde{g}_{eq}$ and $\tilde{g}_{\uparrow\downarrow}$ where saddle points with a certain polarization $P=\left(n_{\uparrow}-n_{\downarrow }\right)/\left(n_{\uparrow}+n_{\downarrow}\right)$ exist and are stable. To this end, we start from a given inverse temperature $\beta$ and polarization $P$. The number equations can be used to calculate the effective chemical potentials $\mu_{\uparrow}^{\prime}$ and $\mu_{\downarrow}^{\prime}$. Subsequently, the conditions $\Tr H\geq 0$ and $\det H\geq 0$ are used to derive the stability condition
\begin{equation}
-\left \vert G_{\uparrow \downarrow }\right \vert \geq G_{eq}\geq -\frac{2}{z}+\sqrt{\frac{4\left( 1-z\right) }{z^{2}}+G_{\uparrow \downarrow }^{2}}
\end{equation}
as a function of the rescaled interaction parameters $G_{eq}=I_{tot}\tilde{g}_{eq}$ and $G_{\uparrow\downarrow}=I_{tot}\tilde{g}_{\uparrow\downarrow}$. Here $I_{tot}$ and $z\in\left[0,1\right]$ (fig. \ref{stabilityfigures}(a) and \ref{stabilityfigures}(b)) are defined as $I_{tot} = I_{\uparrow} +I_{\downarrow} $ and $z=4I_{\uparrow} I_{\downarrow}/I_{tot}^{2}$, where $I_{\sigma}\left(\beta,\mu^{\prime}_{\sigma}\right)=-\left.\partial^{2}\Omega_{sp,kin}/\left(\partial\mu _{\sigma}^{\prime}\right) ^{2}\right\vert_{\beta,\mu_{-\sigma }^{\prime }}$ is a positive function given by
\begin{equation}
I_{\sigma }=\frac{\beta }{2}\int \frac{d^{D}k}{\left( 2\pi \right) ^{D}}\left( \frac{1}{1+\cosh \left[\beta \left( \mathbf{k}^{2}-\mu _{\sigma }^{\prime }\right) \right] }\right) \text{.}
\end{equation}

In fig. \ref{stabilityfigures}(c) the upper and lower boundaries of the stability areas are shown for different values of $z$. Within these boundaries, solutions to the saddle-point equations are stable (minima), provided that they exist. At least one solution exists if a value of $\mu$ can be found which satisfies the equation (\ref{effchempoteq}) that defines the effective chemical potential. For the unpolarized (UP) solutions, the equation becomes $\mu^{\prime}=\mu-\tilde{g}_{eq}/6\pi^{2}$ and a valid solution can always be found by adapting $\mu$.

For the partially polarized (PP) solutions, the saddle points with polarization $P$ only exist at $G_{eq}=G_{\uparrow\downarrow}-6\pi^{2}\zeta^{\prime}I_{tot}/P$ with $\zeta^{\prime}=\left(\mu _{\uparrow}^{\prime}-\mu_{\downarrow}^{\prime}\right)/2$. This represents a straight line in the $\left(G_{\uparrow\downarrow},G_{eq}\right)$-plane, which may intersect the stability area if $0\leq 3\pi ^{2}\zeta ^{\prime}I_{tot}/P\leq 1$. The parameter $3\pi^{2}\zeta ^{\prime}I_{tot}/P$ can be rewritten as the ratio $\chi_{diff}/\chi_{tot}$ between the differential susceptibility $\chi_{diff}=\left.\partial\left(\delta n\right)/\partial\zeta^{\prime}\right\vert_{\beta,\mu^{\prime}}=I_{tot}$ and the total susceptibility $\chi_{tot}=\delta n/\zeta^{\prime }$, which were defined in analogy to the magnetic susceptibility.

\begin{figure}
\includegraphics[scale = 0.8]{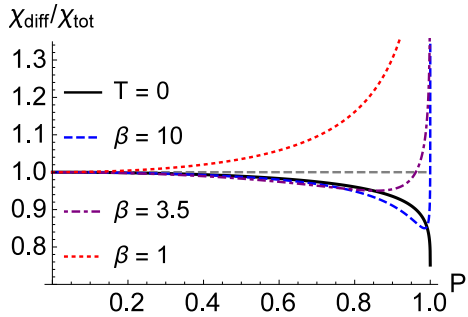}
\caption{\label{susceptibilityfigure}The ratio $\chi_{diff}/\protect\chi_{tot}$ between the differential and total susceptibility as a function of the polarization $P$ for different values of the inverse temperature: $\beta\rightarrow+\infty$ ($T=0$, black solid line), $\beta=10$ (blue dashed line), $\beta=3.5$ (purple dot-dashed line) and $\beta =1$ (red dotted line).}
\end{figure}

In fig. \ref{susceptibilityfigure}, $\chi_{diff}/\chi_{tot}$ is shown for different values of $\beta$. For finite values of $\beta$, $\chi _{diff}/\chi_{tot}\rightarrow+\infty$ if $P\rightarrow 1$. At low temperatures, PP solutions can be stable up to a maximum polarization $P_{\max}$, which decreases as a function of temperature. For $\beta _{\min}\alt 1.715$, the PP saddle points cannot be stable.

Fully polarized (FP) saddle points only exist at temperature zero, because $n_{\sigma }=0$ can only be achieved in the limit $\mu_{\sigma}\rightarrow+\infty $ at non-zero temperatures. This is an important qualitative difference between temperature zero ($\beta\rightarrow+\infty$) and non-zero temperatures (finite values of $\beta$). At zero temperature, FP solutions to the number equations exist for $\zeta ^{\prime}\geq 2^{-1/3}$, which can be induced by the interactions if $G_{eq}\geq G_{\uparrow\downarrow}-3\pi^{2}2^{2/3}I_{tot}$.

\begin{figure}
\includegraphics[width=\linewidth]{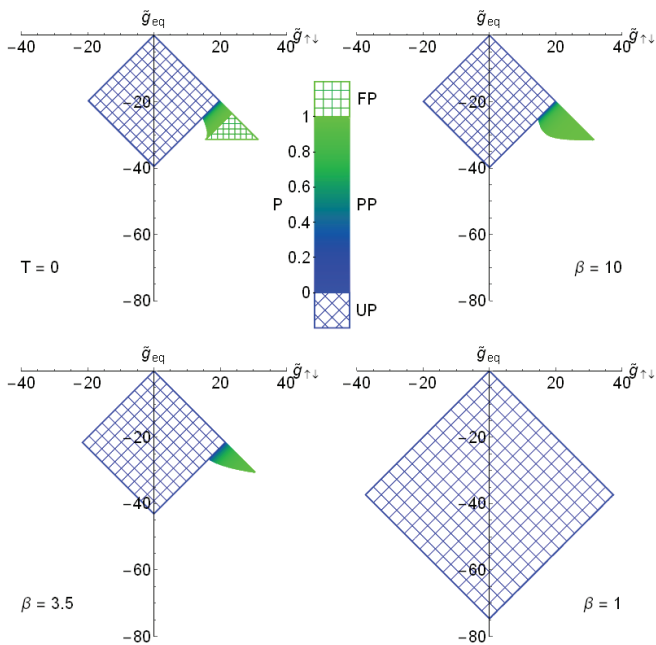}
\caption{\label{phasediagrams}The phase diagram as a function of the modified interaction parameters $\tilde{g}_{eq}$ and $\tilde{g}_{\uparrow\downarrow}$ for different values of the inverse temperature: $\beta\rightarrow+\infty$ ($T=0$), $\beta=10$, $\beta=3.5$ and $\beta=1$. The unpolarized (UP) stability area has blue borders and diagonal blue hatching. The fully polarized (FP) stability area has green borders and darker green horizontal and vertical hatching. The partially polarized (PP) stability area is colored, where the color indicates the polarization.}
\end{figure}

By combining the existence and stability conditions, phase diagrams as a function of the modified interaction parameters $\tilde{g}_{eq}$ and $\tilde{g}_{\uparrow\downarrow}$ are created (fig. \ref{phasediagrams}). They can be related to experimental systems using $a_{s}^{\uparrow\downarrow}k_{F}\approx\tilde{g}_{\uparrow\downarrow}/8\pi$, where $a_{s}^{\uparrow\downarrow}$ is the s-wave scattering length. Stable saddle points are only found in the lowest quadrant ($\tilde{g}_{eq}\leq\left\vert\tilde{g}_{\uparrow \downarrow}\right\vert$). The UP stability area is a square which increases in size as a function of temperature. The PF and PP stability areas shrink and become less polarized as a function of temperature, until they are completely absorbed by the growing UP area. In the PP and FP areas, $a_{s}^{\uparrow\downarrow}k_{F}$ is of order one, which is in agreement with second order predictions for itinerant ferromagnetism in ultracold atomic gases\cite{duineprediction}.

If $\left\vert\tilde{g}_{eq}\right\vert$ or $\left\vert\tilde{g}_{\uparrow\downarrow}\right\vert$ become too large, the system becomes too susceptible to density fluctuations and none of the saddle points can be stable. This stability requirement puts an extra restraint on the Stoner criterion, which only requires $a_{s}^{\uparrow\downarrow}k_{F}$ to be sufficiently large (based on a simple energy argument). This greatly reduces the itinerant ferromagnetic (PP and FP) areas in the phase diagram, which may explain why this state is so hard to find experimentally.  

What happens outside of the stability/existence areas lies outside the scope of this Letter, as in that case the saddle points studied here do not provide a valid description of the system. Possible phases to be considered include superfluid pairing of $\uparrow$ and $\downarrow$ particles, phase separation and non-uniform phases (e.g. density waves in dipolar gases \cite{dipolardensitywave}). The latter two phases may still increase the itinerant ferromagnetic area in the phase diagram.

\textit{Conclusion} In this Letter, we have shown that the fermionic symmetry properties of a pair or correlation of two fermionic particles are not conserved when treating that pair as a (quasi-)boson, as is done when performing the Hubbard-Stratonovich transformation. This effect becomes especially important when studying the exchange interactions for a general interaction potential. We have proposed a correction factor, which includes the effects of exchange in the interaction potential.

We demonstrated this approach in the saddle-point approximation by studying itinerant ferromagnetism in a 3D spin-1/2 Fermi gas. We found that all phases become unstable to density fluctuations when the interactions are sufficiently strong. This greatly limits the area where we expect to find the uniform (PP and FP) itinerant ferromagnetic phases.

\begin{acknowledgments}
The authors would like to thank W. Ketterle and F. Brosens for interesting discussions, and J. P. A. Devreese for a careful reading of the manuscript. E. V. gratefully acknowledges support in the form of a Ph. D. fellowship of the Research Foundation - Flanders (FWO). This work was supported by the following Research Projects of the Research Foundation-Flanders (FWO): G.0119.12N, G.0122.12N, G.0429.15N, and WOG (WO.035.04N). The work was also supported by the Research Fund of the University of Antwerp (J.T.) and ARO (Grant No. W911NF-09-1-0220) (C.A.R.S.dM).
\end{acknowledgments}

\bibliography{ifgeneralinteractions}% Produces the bibliography via BibTeX.

\end{document}